\documentclass[pra,twocolumn,groupedaddress,superscriptaddress,amsmath,amssymb,a4paper]{revtex4}
\usepackage{geometry}
\usepackage{graphicx}
\usepackage{mathrsfs}
\usepackage{amssymb}
\usepackage{amsmath}
\usepackage{amsthm}
\usepackage{amsbsy}
\usepackage{bm}
\usepackage{hyperref}

\theoremstyle{definition}

\newcommand{\bra}[1]{\langle #1|}
\newcommand{\ket}[1]{| #1 \rangle }

\begin{document}

\title{Entanglement robustness in trace decreasing quantum dynamics \\ caused by depolarization and polarization dependent losses}

\author{Sergey N. Filippov}

\affiliation{Steklov Mathematical Institute of Russian Academy of
Sciences, Gubkina St. 8, Moscow 119991, Russia}
\affiliation{Valiev Institute of Physics and Technology of Russian
Academy of Sciences, Nakhimovskii Pr. 34, Moscow 117218, Russia}
\affiliation{Moscow Institute of Physics and Technology,
Institutskii Per. 9, Dolgoprudny, Moscow Region 141700, Russia}

\begin{abstract}
Trace decreasing dynamical maps are as physical as trace
preserving ones; however, they are much less studied. Here we
overview how the quantum Sinkhorn theorem can be successfully
applied to find a two-qubit entangled state which has the
strongest robustness against local noises and losses of quantum
information carriers. We solve a practically relevant problem of
finding an optimal initial encoding to distribute entangled
polarized qubits though communication lines with polarization
dependent losses and extra depolarizing noise. The longest
entanglement lifetime is shown to be attainable with a state that
is not maximally entangled.
\end{abstract}

\maketitle


\section{Introduction} \label{section-introduction}

General physical transformations of quantum states are usually
associated with quantum channels, i.e., completely positive trace
preserving maps (see,
e.g.,~\cite{nielsen-2000,breuer-petruccione-2002,holevo-2012,jagadish-2018}).
However, if we consider a generally nonprojective quantum
measurement, then the induced state transformation is a
\emph{quantum operation}, i.e., a completely positive and trace
nonincreasing map~\cite{davies-lewis-1970,heinosaari-ziman}.

A mathematical condition of complete positivity is equivalent to a
physical condition of positive semidefiniteness for a
composite-system density operator, provided the transformation has
affected a part of the composite system. The complete positivity
condition is long known to be equivalent to positive
semidefiniteness of the so-called Choi matrix~\cite{choi-1975}
(see also~\cite{de-pillis-1967,jamiolkowski-1972}); however, the
same matrix was introduced as a dynamical matrix in an earlier
publication by Sudarshan, Mathews, and Rau~\cite{smr-1961}, where
they implicitly imposed the stronger condition of complete
positivity~\cite[Eq.~(16)]{smr-1961} instead of the weaker
condition of positivity~\cite[Eq.~(12')]{smr-1961}. The trace
preservation condition takes the form of a matrix equality
involving the dynamical matrix~\cite[Eq.~(17)]{smr-1961}, so the
trace-nonincreasing condition takes the form of a matrix
inequality.

Note that a mapping from the space of measurement outcomes to the
set of quantum operations is known in the literature as a quantum
instrument~\cite{holevo-2012,davies-lewis-1970,heinosaari-ziman}.
Repeated applications of the same quantum operation can be used to
simulate non-Hermitian quantum dynamics~\cite{lf-2017,gmsvf-2020},
whereas repeated applications of the same quantum instrument
enable quantum state tomography~\cite{zhuravlev-2020}.
Sequentially intervening open system dynamics with quantum
instruments as in an experiment in Ref.~\cite{xiang-2021} makes it
possible to fully learn a generally non-Markovian quantum
process~\cite{lvgf-2020}.

On the other hand, a loss of quantum information carriers in a
quantum communication line can be viewed as a quantum operation
$\Lambda$ too. In this case, the detection probability ${\rm
tr}\big[ \Lambda[\varrho] \big]$ is the probability to
successfully implement a quantum operation $\Lambda$ for a given
input density operator $\varrho$. If the detection probability is
the same for all initial density operators $\varrho$, then
$\Lambda$ is merely an attenuated quantum channel, i.e., there
exists $0 \leq p \leq 1$ and a quantum channel $\Phi$ such that
$\Lambda = p \Phi$. However, quantum physics is much richer and
there exist such \emph{biased} quantum operations $\Lambda$ that
${\rm tr}\big[ \Lambda[\varrho] \big] \neq {\rm tr}\big[
\Lambda[\varrho'] \big]$ for at least two density operators
$\varrho$ and $\varrho'$~\cite{filippov-2021}. A prominent example
of the latter case is polarization dependent
losses~\cite{filippov-2021,gisin-1997}, for which horizontally and
vertically polarized photons have different loss probabilities.
That asymmetry significantly affects the way one should encode
quantum information to reliably transmit quantum information
through such a lossy channel, namely, special codes with entangled
states are shown to perform better than codes with disentangled
states~\cite{filippov-2021}.

A one-parameter family of quantum operations $\{\Lambda(t)\}_{t
\geq 0}$ represents a process of physical evolution in time $t
\geq 0$. If $\varrho(0)$ is an initial density operator of the
system, then $\Lambda(t)[\varrho(0)]$ is a subnormalized density
operator at time $t$. In a general physical process, the detection
probability ${\rm tr}\big[ \Lambda(t)[\varrho(0)] \big]$ does not
have to decrease monotonically as quantum information carriers can
potentially return to the communication
line~\cite{jin-2015,mataloni-2019}. Experimentalists usually
postselect successful realizations (e.g., in biphoton
production~\cite{kulik-2006}) and study dynamics of the
conditional output states
\begin{equation} \label{conditional-output}
\widetilde{\varrho}(t) := \frac{\Lambda(t)[\varrho(0)]}{{\rm tr}
\big[ \Lambda(t)[\varrho(0)] \big]}.
\end{equation}

\noindent Note that the map $\varrho(0) \rightarrow
\widetilde{\varrho}(t)$ in nonlinear, so focusing solely on the
dynamics of $\widetilde{\varrho}(t)$ may lead to a
misidentification of non-Markovianity of
$\Lambda(t)$~\cite{filippov-QP-2021}.

We consider a physically relevant problem of two-qubit
entanglement distribution through lossy communication
lines~\cite{filippov-jms-2019,filippov-jms-2021}. Let
$\varrho_{12}(0)$ be an initial density operator and $\Lambda(t) =
\Lambda_1(t) \otimes \Lambda_2(t)$, where $\Lambda_i(t)$ is a
qubit operation describing loss and noise accumulated in the
$i$-th communication line by time $t$. The goal of this paper is
to present a recipe on how one should prepare an initial entangled
state $\varrho_{12}(0)$ so that $\widetilde{\varrho}_{12}(t)$
remains entangled during the trace decreasing dynamics for the
longest possible time. Entanglement represents a useful resource
for quantum communication and device-independent key
distribution~\cite{xu-2013}, so its preservation is of high
importance to applications. In this paper, we are interested in
the fundamental separation between entangled and disentangled
states and pay no attention to the ``amount'' of entanglement. We
do that because an arbitrary little but nonzero two-qubit
entanglement can potentially be purified: many copies of poorly
entangled states can be transformed into a less number of almost
maximally entangled two-qubit states~\cite{horodecki-1997}. In
view of this, we focus on the maximum permissible noise level,
exceeding which no entanglement-enabled protocol is feasible and
below which any entanglement-enabled protocol is fundamentally
possible. An entanglement lifetime (a disentangling time, an
entanglement sudden death time~\cite{almeida-2007}) of an initial
state $\varrho_{12}(0)$ is defined as the minimal time $\tau$ such
that $\widetilde{\varrho}_{12}(t)$ is disentangled for all $t \geq
\tau$. The maximal possible entanglement lifetime
\begin{equation}
\widetilde{\tau} = \max_{\varrho_{12}(0)} \tau
\end{equation}

\noindent provides the fundamental restriction on the length of
quantum communications lines for entanglement distribution.

Our study follows a similar analysis made for trace preserving
maps~\cite{frz-2012,ffk-2018}; however, the trace decreasing
nature of $\Lambda(t)$ modifies the result. If $\Lambda(t)$ is
biased, then some states have higher detection probability than
others, which increases their contribution to
Eq.~\eqref{conditional-output}. The optimal state
$\varrho_{12}^{\rm opt}(0)$ makes allowance for that effect and is
not maximally entangled in general. A technique to solve the
optimization problem posed is based on the quantum Sinkhorn
theorem~\cite{gurvits-2004,georgiou-2015,aubrun-2015,aubrun-2017}
that also finds applications in the study of quantum channel
capacity~\cite{filippov-2018}. The quantum Sinkhorn theorem
relates strictly positive quantum maps with unital ones and
enables us to use the known results on entanglement robustness
against unital quantum noises~\cite{frz-2012,ffk-2018}. We
implement that research programme in
Section~\ref{section-ent-Sinkhorn}.

In addition to a general result, in Section~\ref{section-pdl}, we
elaborate the case of polarization dependent losses accompanied by
depolarization. This model of loss and noise describes effects in
some optical fibers and attracts increasing attention in the
literature~\cite{kirby-2019,li-2018}. In this model, each
$\Lambda_i(t)$ is defined by three parameters: the depolarization
rate $\gamma$ and the attenuation coefficients for horizontally
and vertically polarized photons, $\gamma_H$ and $\gamma_V$. We
develop a technique on how to find the optimal state
$\varrho_{12}^{\rm opt}(0)$ and the maximal entanglement lifetime
provided the above parameters are known for both lines
$\Lambda_1(t)$ and $\Lambda_2(t)$.

\section{Entanglement dynamics and quantum Sinkhorn theorem}
\label{section-ent-Sinkhorn}

A bipartite density operator $\varrho_{12}$ on a finite
dimensional Hilbert space ${\cal H}_1 \otimes {\cal H}_2$ is
called disentangled with respect to the bipartition $1|2$ if it
adopts a convex sum representation of the
form~\cite{werner-1989,horodecki-2009}
\begin{equation} \label{entangled}
\varrho_{12} = \sum_k p_k \varrho_1^{(k)} \otimes \varrho_2^{(k)},
\end{equation}

\noindent where $\{p_k\}_k$ is a probability distribution and
$\{\varrho_i^{(k)}\}_k$ is a collection of density operators on
${\cal H}_i$, $i=1,2$. A subnormalized density operator
$\varrho_{12}$ with ${\rm tr}[\varrho_{12}] \leq 1$ is
disentangled if Eq.~\eqref{entangled} represents a conic sum,
i.e., $p_k \geq 0$ for all $k$.

Let $\{\ket{0},\ket{1}\}$ be a standard qubit basis such that
$\sigma_x = \ket{0}\bra{1} + \ket{1}\bra{0}$, $\sigma_y = -i
\ket{0}\bra{1} + i \ket{1}\bra{0}$, and $\sigma_z = \ket{0}\bra{0}
- \ket{1}\bra{1}$ are the conventional Pauli operators. The
maximally entangled two-qubit state reads
$\ket{\psi_+}\bra{\psi_+}$, where $\ket{\psi_+} =
\frac{1}{\sqrt{2}}(\ket{00}+\ket{11})$. Note that a reduced
density operator of either qubit from the maximally entangled
two-qubit pair is in the maximally mixed state $\frac{1}{2}I$,
where $I = \ket{0}\bra{0} + \ket{1}\bra{1}$ is the identity
operator.

Let us consider such qubit transformations for which the maximally
mixed state $\frac{1}{2}I$ is a fixed point. A linear qubit map
$\Upsilon$ is called unital if $\Upsilon[I] = I$. If $\Upsilon$ is
a unital quantum channel, then the von Neumann entropy of the
output state $\Upsilon[\varrho]$ is not less than the von Neumann
entropy of the input state $\varrho$ for any density operator
$\varrho$.

Two-qubit entanglement dynamics in presence of identical local
unital noises of the form $\Upsilon \otimes \Upsilon$ is studied
in Ref.~\cite{frz-2012}. The maximally entangled state
$\ket{\psi_+}\bra{\psi_+}$ is the most robust against local unital
noises in the sense that $\Upsilon \otimes \Upsilon
[\varrho_{12}]$ is disentangled for any density operator
$\varrho_{12}$ whenever $\Upsilon \otimes \Upsilon
[\ket{\psi_+}\bra{\psi_+}]$ is disentangled~\cite{frz-2012}.

A generalization of that result for different unital local noises
of the form $\Upsilon \otimes \Upsilon'$ is obtained in
Ref.~\cite{ffk-2018}, where a state with the ultimate entanglement
robustness is shown to be the maximally entangled state of the
form
\begin{equation} \label{robust-unital}
\ket{\psi_{\Upsilon\otimes\Upsilon'}} = \frac{1}{\sqrt{2}}\left(
\ket{\varphi} \otimes \ket{\chi} + \ket{\varphi_{\perp}} \otimes
\ket{\chi_{\perp}} \right),
\end{equation}

\noindent with $\{\ket{\varphi},\ket{\varphi_{\perp}}\}$
($\{\ket{\chi},\ket{\chi_{\perp}}\}$) being orthogonal
eigenvectors of some traceless eigenoperator of $\Upsilon$
($\Upsilon'$).

If both $\Upsilon$ and $\Upsilon'$ are diagonal in the basis of
Pauli operators $\sigma_x$, $\sigma_y$, $\sigma_z$, i.e.,
\begin{eqnarray}
&& \Upsilon[\varrho] = \frac{1}{2} {\rm tr}[\varrho] I +
\frac{1}{2}
\sum_{k=x,y,z} \lambda_k {\rm tr}[\sigma_k \varrho] \sigma_k, \label{unital-parameters} \\
&& \Upsilon'[\varrho] = \frac{1}{2} {\rm tr}[\varrho] I +
\frac{1}{2} \sum_{k=x,y,z} \lambda'_k {\rm tr}[\sigma_k \varrho]
\sigma_k, \label{unital-parameters-prime}
\end{eqnarray}

\noindent and additionally $\lambda_x \geq \lambda_y \geq
\lambda_z \geq 0$, $\lambda_x' \geq \lambda_y' \geq \lambda_z'
\geq 0$, then $\ket{\psi_{\Upsilon\otimes\Upsilon'}} =
\ket{\psi_+}$.

If parameters $\lambda_k$ and $\lambda'_k$ in
Eqs.~\eqref{unital-parameters} and \eqref{unital-parameters-prime}
are functions of time, then we deal with a local unital dynamics
$\Upsilon(t) \otimes \Upsilon'(t)$ which preserves entanglement of
the maximally entangled state as long as~\cite{ffk-2018}
\begin{equation} \label{equation-lifetime}
\lambda_x(t) \lambda_x'(t) + \lambda_y(t) \lambda_y'(t) +
\lambda_z(t) \lambda_z'(t) > 1.
\end{equation}

A linear map $\Lambda$ on operators in ${\cal H}$ is called
strictly positive if $\Lambda[\varrho]$ is positive definite for
all nonzero positive semidefinite operators $\varrho$. Strictly
positive maps belong to the interior of the cone of positivity
preserving maps~\cite{aubrun-2015} and are also referred to as
positivity improving ones~\cite{georgiou-2015}. If $\Lambda$ is
strictly positive, then by Proposition 2.32 in~\cite{aubrun-2017}
there exist positive definite operators $A$ and $B$ such that the
map
\begin{equation} \label{Sinkhorn}
\Upsilon = \Phi_A \circ \Lambda \circ \Phi_B
\end{equation}

\noindent is trace preserving and unital. Here, $\Phi_X[\varrho] =
X \varrho X^{\dag}$ and $\circ$ denotes the map concatenation. If
$\Lambda$ is completely positive and strictly positive, then
$\Upsilon$ is a unital quantum channel. The relation
\eqref{Sinkhorn} is known as the quantum Sinkhorn theorem
originally discovered in Ref.~\cite{gurvits-2004}, rediscovered
for positivity improving completely positive maps in
Ref.~\cite{georgiou-2015}, and finally clarified in
Refs.~\cite{aubrun-2015,aubrun-2017}. One can express the
operators $A$ and $B$ through $A = \sqrt{S}$ and $B =
(\Lambda^{\dag}[S])^{-1/2}$, where a positive definite operator
$S$ is a fixed point of the map $F[S] =
(\Lambda[(\Lambda^{\dag}[S])^{-1}])^{-1}$, where $\Lambda^{\dag}$
denotes a dual linear map with respect to $\Lambda$, i.e., ${\rm
tr} \big[ \Lambda^{\dag}[X] Y \big] = {\rm tr} \big[ X \Lambda[Y]
\big]$ for all $X,Y$. A methodology to explicitly find operators
$A$ and $B$ for trace preserving qubit maps $\Lambda$ is given in
Ref.~\cite{ffk-2018}; however, a methodology for trace decreasing
maps is still missing. In Section~\ref{section-pdl}, we partially
fill this gap for a physically relevant scenario of combined noise
and loss.

The inverse relation to Eq.~\eqref{Sinkhorn} reads
\begin{equation} \label{Sinkhorn-inverse}
\Lambda = \Phi_{A^{-1}} \circ \Upsilon \circ \Phi_{B^{-1}}
\end{equation}

\noindent and enables us to find the structure of the optimal
state $\varrho^{\rm opt}(0)$. Indeed, since the map $\Phi_X$ has a
single Kraus operator $X$, the operator $\Lambda \otimes
\Lambda'[\varrho(0)]$ is entangled if and only if $(\Upsilon
\otimes \Upsilon') \circ (\Phi_{B^{-1}} \otimes \Phi_{B'^{-1}})
[\varrho(0)]$ is entangled. On the other hand, the most robust
entangled state against the noise $\Upsilon \otimes \Upsilon'$ is
$\ket{\psi_{\Upsilon\otimes\Upsilon'}}\bra{\psi_{\Upsilon\otimes\Upsilon'}}$
given by Eq.~\eqref{robust-unital}, so $(\Phi_{B^{-1}} \otimes
\Phi_{B'^{-1}}) [\varrho(0)] \propto
\ket{\psi_{\Upsilon\otimes\Upsilon'}}\bra{\psi_{\Upsilon\otimes\Upsilon'}}$.
Inverting the map $\Phi_{B^{-1}} \otimes \Phi_{B'^{-1}}$, we get
$\varrho^{\rm opt}(0) =
\ket{\psi_{\Lambda\otimes\Lambda'}}\bra{\psi_{\Lambda\otimes\Lambda'}}$,
where
\begin{equation} \label{robust-general}
\ket{\psi_{\Lambda\otimes\Lambda'}} = \frac{B(\widetilde{\tau})
\otimes B'(\widetilde{\tau})
\ket{\psi_{\Upsilon\otimes\Upsilon'}}}{\sqrt{\bra{\psi_{\Upsilon\otimes\Upsilon'}}
B^{\dag}(\widetilde{\tau})B(\widetilde{\tau}) \otimes
B'(\widetilde{\tau})^{\dag}B'(\widetilde{\tau})
\ket{\psi_{\Upsilon\otimes\Upsilon'}}}}
\end{equation}

\noindent and $\widetilde{\tau}$ is the maximal entanglement
lifetime under noise $\Upsilon(t) \otimes \Upsilon'(t)$ determined
in~\cite[Proposition~1]{ffk-2018}. In the case when both
$\lambda_x(t) \geq \lambda_y(t) \geq \lambda_z(t) \geq 0$ and
$\lambda_x'(t) \geq \lambda_y'(t) \geq \lambda_z'(t) \geq 0$, one
can substitute $\ket{\psi_+}$ for
$\ket{\psi_{\Upsilon\otimes\Upsilon'}}$ and find the maximal
entanglement lifetime $\widetilde{\tau}$ as the smallest $t > 0$
for which the inequality~\eqref{equation-lifetime} is violated.

\section{Entanglement distribution in presence of depolarization and polarization dependent
losses} \label{section-pdl}

The effect of polarization dependent losses is that the states
with different polarization are attenuated
differently~\cite{gisin-1997,filippov-2021,kirby-2019,li-2018}.
Let the horizontally and vertically polarized states be those that
are the least and most attenuated, or vice versa. By $\gamma_H$
and $\gamma_V$ denote the attenuation rates for horizontally and
vertically polarized photons, respectively. Let
$(\ket{H},\ket{V})$ be a conventional basis composed of the
horizontally and vertically polarized one-photon states. In what
follows, we will identify the basis $(\ket{H},\ket{V})$ with the
standard basis $(\ket{0},\ket{1})$ used in
Section~\ref{section-ent-Sinkhorn}. Then the combined effect of
polarization dependent losses and depolarization with the rate
$\gamma$ on a polarization qubit is described the following master
equation:
\begin{eqnarray}
\frac{d \varrho(t)}{dt} &=& - \frac{1}{2} \Big\{ \gamma_H
\ket{H}\bra{H} + \gamma_V \ket{V}\bra{V},\varrho(t) \Big\}
\nonumber\\
&& + \frac{\gamma}{4}\sum_{k=x,y,z} \big( \sigma_k \varrho(t)
\sigma_k - \varrho(t) \big), \label{master-pdl-depolarization}
\end{eqnarray}

\noindent where $\{ \cdot, \cdot \}$ stands for the
anticommutator. Eq.~\eqref{master-pdl-depolarization} defines the
dynamical semigroup $\Lambda(t)$ that is trace decreasing if
$\gamma_H > 0$ and $\gamma_V > 0$.

Let us consider a matrix representation (see,
e.g.,~\cite{heinosaari-ziman}) of the qubit map $\Lambda(t)$,
i.e., a $4 \times 4$ matrix $M(t)$ whose elements are defined
through
\begin{equation}
M_{ij}(t) = \frac{1}{2} {\rm tr} \big[ \sigma_i
\Lambda(t)[\sigma_j] \big], \quad i,j = 0,x,y,z,
\end{equation}

\noindent where $\sigma_0 = I$. Some tedious yet straightforward
algebra yields
\begin{eqnarray}
&& M(t) = \left(%
\begin{array}{cccc}
  a(t) & 0 & 0 & b(t) \\
  0 & c(t) & 0 & 0 \\
  0 & 0 & c(t) & 0 \\
  b(t) & 0 & 0 & d(t) \\
\end{array}%
\right), \label{matrix-representation} \\
&& a(t) = e^{-\frac{1}{2}(\gamma + \gamma_H + \gamma_V) t} \bigg( \cosh \frac{\sqrt{\gamma^2+(\gamma_H - \gamma_V)^2} \, t}{2}  \nonumber\\
&& \quad + \frac{\gamma}{\sqrt{\gamma^2+(\gamma_H - \gamma_V)^2}} \sinh \frac{\sqrt{\gamma^2+(\gamma_H - \gamma_V)^2} \, t}{2} \bigg), \qquad \label{a-t} \\
&& b(t) = - \frac{\gamma_H - \gamma_V}{\sqrt{\gamma^2+(\gamma_H - \gamma_V)^2}} e^{-\frac{1}{2}(\gamma + \gamma_H + \gamma_V) t} \nonumber\\
&& \quad \times \sinh \frac{\sqrt{\gamma^2+(\gamma_H - \gamma_V)^2} \, t}{2} ,\\
&& c(t) = e^{-\frac{1}{2}(2\gamma + \gamma_H + \gamma_V) t}, \\
&& d(t) = e^{-\frac{1}{2}(\gamma + \gamma_H + \gamma_V) t} \bigg( \cosh \frac{\sqrt{\gamma^2+(\gamma_H - \gamma_V)^2} \, t}{2}  \nonumber\\
&& \quad - \frac{\gamma}{\sqrt{\gamma^2+(\gamma_H - \gamma_V)^2}}
\sinh \frac{\sqrt{\gamma^2+(\gamma_H - \gamma_V)^2} \, t}{2}
\bigg). \qquad \label{d-t}
\end{eqnarray}

To apply the quantum Sinkhorn theorem to a map $\Lambda(t)$ with
the matrix representation~\eqref{matrix-representation} we need to
find a fixed point of the map $F[S]$, i.e., to solve a matrix
equation
\begin{equation} \label{fixed-point-eq}
S = \Bigg( \Lambda(t)\bigg[\left(\Lambda(t)^{\dag}[S]\right)^{-1}
\bigg] \Bigg)^{-1}.
\end{equation}

\noindent Note that in our case $\Lambda(t)^{\dag} = \Lambda(t)$.
Due to the phase covariance of $\Lambda(t)$~\cite{fgl-2020}, we
seek the operator $S$ in the form of an ansatz $S = I + s
\sigma_z$. If $b(t) \neq 0$, this results in the following
solution:
\begin{equation}
s(t) = - \frac{a(t)+d(t) - \sqrt{[a(t)+d(t)]^2 - 4
b^2(t)}}{2b(t)}.
\end{equation}

\noindent Note that $S = I + s \sigma_z$ is positive definite if
$a(t) + d(t) \geq 2 |b(t)| \neq 0$ which is automatically
fulfilled for expressions~\eqref{a-t}--\eqref{d-t} if $t>0$.
Substituting $A = \sqrt{S}$ and $B =
(\Lambda(t)^{\dag}[S])^{-1/2}$ into Eq.~\eqref{Sinkhorn}, we get
the corresponding unital map $\Upsilon(t)$ that has the
form~\eqref{unital-parameters} with parameters
\begin{eqnarray}
&& \lambda_x(t) = \lambda_y(t) = \frac{2 c(t)}{a(t)-d(t)+\sqrt{[a(t)+d(t)]^2 - 4 b^2(t)}}, \nonumber\\
&& \\
&& \lambda_z(t) = \frac{4[a(t)d(t)-b^2(t)]}{\left\{
a(t)-d(t)+\sqrt{[a(t)+d(t)]^2 - 4 b^2(t)} \right\}^2}.
\end{eqnarray}

\noindent The obtained expressions satisfy the conditions
$\lambda_x(t) \geq \lambda_y(t) \geq \lambda_z(t) \geq 0$, so we
can use simplified results at the end of
Section~\ref{section-ent-Sinkhorn}.

Given two communication lines $\Lambda(t)$ and $\Lambda'(t)$ with
parameters $\gamma_H$, $\gamma_V$, $\gamma$ and $\gamma'_H$,
$\gamma'_V$, $\gamma'$, respectively, the maximal entanglement
lifetime $\widetilde{\tau}$ is a solution of the equation
\begin{equation} \label{equation-lifetime-pdl}
\lambda_x(\widetilde{\tau}) \lambda'_x(\widetilde{\tau}) +
\lambda_y(\widetilde{\tau}) \lambda'_y(\widetilde{\tau}) +
\lambda_z(\widetilde{\tau}) \lambda'_z(\widetilde{\tau}) = 1.
\end{equation}

\noindent The optimal state $\varrho^{\rm opt}(0) =
\ket{\psi_{\Lambda\otimes\Lambda'}}\bra{\psi_{\Lambda\otimes\Lambda'}}$
is defined by the normalized vector
\begin{equation} \label{optimal-pdl}
\ket{\psi_{\Lambda\otimes\Lambda'}} \propto B(\widetilde{\tau})
\otimes B'(\widetilde{\tau}) (\ket{HH} + \ket{VV}),
\end{equation}

\noindent where
\begin{eqnarray}
B(\widetilde{\tau}) & = &
\frac{\ket{H}\bra{H}}{\sqrt{a(\widetilde{\tau})+b(\widetilde{\tau})+s(\widetilde{\tau})[b(\widetilde{\tau})+d(\widetilde{\tau})]}}
\nonumber\\
&& +
\frac{\ket{V}\bra{V}}{\sqrt{a(\widetilde{\tau})-b(\widetilde{\tau})+s(\widetilde{\tau})[b(\widetilde{\tau})-d(\widetilde{\tau})]}}
\label{B-optimal}
\end{eqnarray}

\noindent and $B'(\widetilde{\tau})$ is obtained from
Eq.~\eqref{B-optimal} by replacing $a \rightarrow a'$, $b
\rightarrow b'$, $c \rightarrow c'$, $d \rightarrow d'$.

\begin{figure}
\includegraphics[width=8cm]{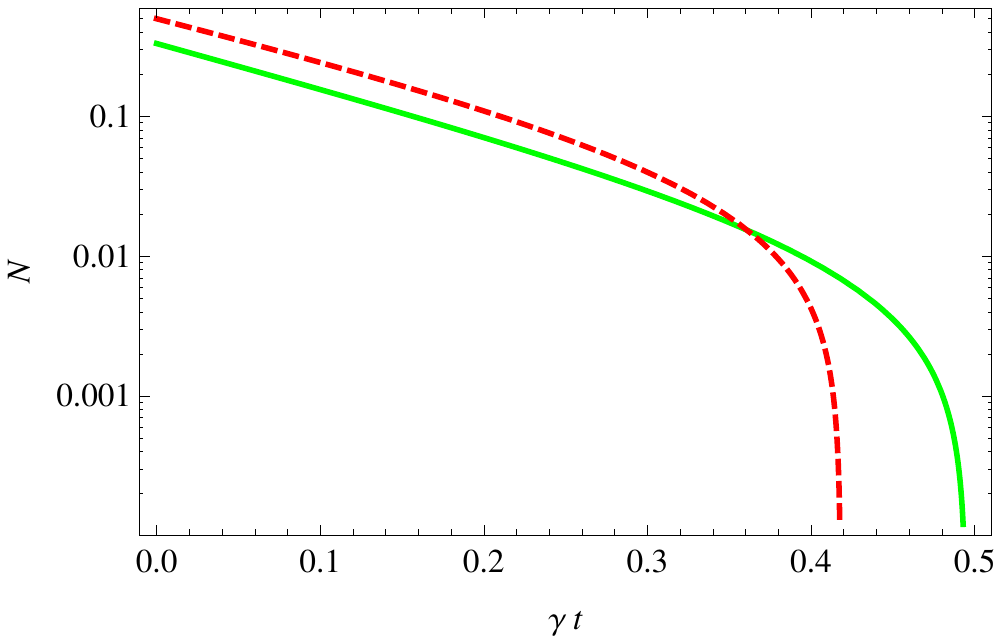}
\caption{\label{figure} Effect of local depolarization and
polarization dependent losses, $\Lambda(t) \otimes \Lambda(t)$
with $\gamma_H = \gamma$ and $\gamma_V = 5 \gamma$, on two-qubit
entanglement dynamics (negativity) for different initial states:
the maximally entangled one (red dashed line) and the optimal one
(green solid line).}
\end{figure}

Despite the fact that the final expression for the optimal
state~\eqref{optimal-pdl} is rather involved, it is analytically
derived and can be further explored. In Fig.~\ref{figure} we
depict the evolution of the entanglement quantifier $N$ called
negativity~\cite{horodecki-2009,zyczkowski-1998} for the maximally
entangled initial state $\ket{\psi_+}\bra{\psi_+}$ and the optimal
state~\eqref{optimal-pdl}. If $\varrho$ is a density operator of a
bipartite system, then $N(\varrho) = \frac{1}{2}(\|({\rm Id}
\otimes T)[\varrho]\|_1 - 1)$, where ${\rm Id}$ is the identity
transformation, $T$ is the transposition in the standard basis,
and $\|X\|_1 = {\rm tr}\sqrt{X^{\dag}X}$. A two qubit state
$\varrho$ is entangled if and only if $N(\varrho)
> 0$~\cite{horodecki-1996}. Fig.~\ref{figure} shows that although the state $\Lambda(t) \otimes
\Lambda'(t)[\varrho^{\rm opt}(0)]$ exhibits less entanglement in
the beginning of evolution as compared to $\Lambda(t) \otimes
\Lambda'(t)[\ket{\psi_+}\bra{\psi_+}]$, the inverse relation takes
place after some time. Finally, after some time $\Lambda(t)
\otimes \Lambda'(t)[\ket{\psi_+}\bra{\psi_+}]$ becomes separable
whereas $\Lambda(t) \otimes \Lambda'(t)[\varrho^{\rm opt}(0)]$
remains entangled. This phenomenon illustrates how a less
entangled optimal state outperforms the maximally entangled state
in long term.

\section{Conclusions} \label{section-conclusions}

Using a general technique of the quantum Sinkhorn theorem and the
previously known results for trace preserving quantum maps, we
have studied the two-qubit entanglement lifetime under arbitrary
trace decreasing local operations
(Section~\ref{section-ent-Sinkhorn}). An analytical treatment of
the quantum Sinkhorn theorem for a particular quantum map
$\Lambda(t)$ is a challenge because Eq.~\eqref{fixed-point-eq} is
rather difficult to resolve. Nevertheless, we have managed to
derive the explicit form of the Sinkhorn theorem for trace
decreasing qubit operations with the matrix
representation~\eqref{matrix-representation}. This analytical
result advances our understanding of the quantum Sinkhorn theorem
beyond the trace-preserving case of phase-covariant qubit maps,
which was the only non-trivial class of maps with explicit
analytical expressions for $\Upsilon$, $A$, and
$B$~\cite{filippov-2018,ffk-2018} known before this work.

We applied the proposed approach to the analysis of entanglement
dynamics of polarization-encoded two-qubit states subjected to the
combined effect of depolarization and polarization dependent
losses (Section~\ref{section-pdl}). Our goal was to find the
longest entanglement lifetime $\widetilde{\tau}$ among all
possible initial states. The longest entanglement lifetime
determines a fundamental noise level exceeding which no
entanglement-enabled protocol can be implemented. We expressed
$\widetilde{\tau}$ as a solution of the analytically derived
Eq.~\eqref{equation-lifetime-pdl}. The optimal initial
state~\eqref{optimal-pdl}---that exhibits the strongest robustness
against depolarization and polarization dependent losses---is not
maximally entangled if $\gamma_H \neq \gamma_V$. The optimal state
makes allowance for the difference in attenuation coefficients
($\gamma_H $ and $\gamma_V$) and has a higher contribution of
those polarization component, which decays more rapidly.

\begin{acknowledgements}
The study in Sections~\ref{section-ent-Sinkhorn},
\ref{section-pdl}, and~\ref{section-conclusions} was supported by
the Russian Science Foundation under Project No. 19-11-00086 and
performed in Steklov Mathematical Institute of Russian Academy of
Sciences. The review in Section~\ref{section-introduction} was
written in Moscow Institute of Physics and Technology as well as
in Valiev Institute of Physics and Technology of Russian Academy
of Sciences, where the author was supported by Program No.
0066-2019-0005 of the Russian Ministry of Science and Higher
Education.
\end{acknowledgements}

\end{document}